\documentclass[12pt]{amsart}

\usepackage{amsmath,amssymb,enumerate}

\usepackage{epsfig,fancyhdr,color}%,showkeys,amsmidx

\usepackage{amssymb}
\usepackage{amsmath,amsthm}
\usepackage{bbm}
\usepackage{latexsym}
\usepackage{amscd}
\usepackage{psfrag}
\usepackage{graphicx}
\usepackage[all]{xy}

\usepackage{rotating}
\usepackage{adjustbox}

\usepackage[font=small,labelfont=bf]{caption}

\usepackage{mathtools}

\usepackage{tikz-cd}

\usepackage{stackrel}

\usepackage{mathabx,epsfig}

\usepackage[utf8]{inputenc}

\numberwithin{equation}{section}

\definecolor{NoteColor}{rgb}{1,0,0}

\newtheorem*{theorem 1}{\rm\bf Proposition 1}
\newtheorem*{theorem 2}{\rm\bf Proposition 2}

\theoremstyle{definition}

\theoremstyle{remark}

\def\interieur#1{\mathord{\mathop{\kern 0pt #1}\limits^\circ}}

\begin{document}

\title[Protein Geometry, Function and Mutation]{Protein Geometry, Function and Mutation}
\
\author{Robert Penner}
\address {\hskip -2.5ex Institut des Hautes \'Etudes Scientifiques\\
35 route des Chartres\\
Le Bois Marie\\
91440 Bures-sur-Yvette\\
France\\
{\rm and}~Mathematics Department,
UCLA\\
Los Angeles, CA 90095\\USA}
\email{rpenner{\char'100}ihes.fr}

\thanks{It is a pleasure to thank Minus van Baalen, Misha Gromov, Pablo Guardado-Calvo,  Konstantin Khrapko and Nadya Morozova for helpful discussions.}

 \date{\today}

% ---------------------------------------

\begin{abstract}
This survey for mathematicians summarizes several works by the author on protein geometry and protein function with applications to viral glycoproteins in general and the spike glycoprotein of the SARS-CoV-2 virus in particular.  Background biology and biophysics are sketched.
This body of work culminates in a postulate that protein secondary structure regulates mutation, with backbone hydrogen bonds  materializing in critical regions to avoid mutation, and disappearing from other regions to enable it.
%This appears to be the case for the spike glycoprotein of SARS-CoV-2.
\end{abstract}

\maketitle

%\noindent \rule{0.165\textwidth}{0.4pt}

{\let\thefootnote\relax}
\setcounter{footnote}{0} 

\subsection{Introduction}
This survey summarizes a series of papers \cite{pennerjcb1,pennerjcb2,pennercmb,pennervac} by the author inferring protein function and mutation from protein geometry, together with the applications to date.
%An earlier such survey with non-zero intersection here is \cite{pennerbull}.  
%Corrections and improvements to the published works are directly presented here and retrospectively detailed in a closing section of corrigenda.
Continuing to learn basic biology, the author increasingly finds geometry often at the heart of the matter, from enzymes to viruses to immunology to cell signaling to morphogenesis and beyond. 
For reasons to be explained, viral glycoproteins provided the first proving ground for the new geometric techniques, coincidentally predating the pandemic, with attention naturally subsequently focused on the SARS-CoV-2 spike glycoprotein in particular, to be called here simply the {\sl spike}.

The method to be explained depends upon a database of protein geometry computed in the the joint work \cite{naturepaper}, whose application described here provides a novel tool in structural biology.  This tool is employed to predict protein function in \cite{pennerjcb1}, as implemented for the spike in \cite{pennerjcb2,pennercmb}, and to explain a general principle, first articulated in \cite{pennervac}, of protein mutation and evolution based on the spike.  The current paper summarizes the intellectual arc of all these works.  Only as much of the background biology and biophysics will be explained as is necessary.  Recommended standard textbooks are: chemistry \cite{pauling}; molecular cell biology \cite{watson}; protein biophysics \cite{AFbook};  virology  \cite{viruses};
and immunology \cite{Janeway}.  

The 20 gene-encoded amino acids combine (as described in $\S$\ref{sec:chem}) to form proteins. Proteins are expressed in organisms as determined by the chromosomal DNA, then transcribed to RNA, which is spliced and combined by cellular mechanisms, then translated to protein, which is finally decorated with sugars and specific chemical modifications.
% by further cellular processes.  

All four bio-macromolecules, namely DNA, RNA, protein and sugar, interact along with other chemical compounds, to define the biological activity of cells.  Proteins are the scaffolding, instigators and inhibitors along with RNA, and most especially workhorses of these activities.  As such, a protein must reliably adopt its own characteristic shape necessary for biological function, an {\sl aperiodic crystal} as Erwin Schr\"odinger termed it, or a {\sl key-lock mechanism} according to Emil Fischer.  

The cellular construct that transcribes DNA to RNA is an enzyme called {\sl polymerase}, which effectively crawls along the mature DNA, sequentially producing the nascent RNA by
appending its consecutive nucleic acid residues, called {\sl bases}, one at a time.
The cellular construct that translates messenger RNA (mRNA) to protein is called the {\sl ribosome}, itself comprised of proteins and RNA.  It crawls along the mature mRNA reading its bases three at a time in order to determine one amino acid, which is added along the extending nascent protein.  The {\sl genetic code} determines which amino acid  to append  from a triplet of bases.

There are already exceptions to what has been stated since there are actually 20 {\sl standard} gene-encoded
amino acids, one of which is actually an imino acid, plus 2 other uncommon amino acids more recently discovered, as well as yet another occurring by chemical modification on one of the 20.  There are thus in fact 20+2+1 amino acids in our current understanding. 

Moreover, this notion of an aperiodic crystal is also a bit misleading since thermal fluctuations cause a protein to wiggle about overall, especially in its so-called {\sl intrinsically disordered domains}, while negligibly fluctuating at specific locations in the key-lock. Even the genetic code has rare exceptions in cellular organelles called mitochondria.

These provisos illustrate The First Law of Mathematical Biology, which states that there is only one such law, or in other words, in biology every statement except this one has exceptions.  There are thus only statements which are usually or generically true. There are no theorems.

\subsection{Chemistry}\label{sec:chem}

Let C, H, N and O respectively denote a Carbon, Hydrogen, Nitrogen and Oxygen atom.  An {\sl amino acid} is an organic compound containing one amine NH$_2$
and one carboxyl CO$_2$H functional group, each covalently bonded to a so-called {\sl residue} R specific to the amino acid.  These serially combine in consecutive pairs condensing off a water molecule and creating a {\sl peptide bond}
C=N from the C in one carboxyl to the N in the next amine to produce a linear polymer.  
Letting C$^\alpha$ denote the $\alpha^{th}$ carbon
of a residue (the ``first'' carbon), there is a resulting {\sl backbone}: C$^\alpha$ -- C = N -- C$^\alpha$ -- C = N -- $\cdots$ -- C$^\alpha$.  The backbone has its canonical orientation
from the remaining amine end, the {\sl N terminus}, to the remaining carboxyl end, the {\sl C terminus}

Four consecutive backbone atoms containing a peptide bond, together with the O and H attached to the backbone respectively remaining from the carboxyl and amine after water condensation, comprise
a {\sl peptide group}
$\begin{smallmatrix}
&\\
&&&{\rm O}&&&&&&{\rm C}^\alpha\cr
&&&&\begin{turn}{-45}=\end{turn}&&&&\begin{turn}{45}--\end{turn}\cr
&&&&&{\rm C}&=&{\rm N}\cr
&&&&\begin{turn}{45}--\end{turn}&&&&\begin{turn}{-45}--\end{turn}\cr
&&&{\rm C}^\alpha&&&&&&{\rm H}\cr
&\\
\end{smallmatrix}$. It is a consequence of quantum chemistry, namely the {\sl ${\rm sp}^2-{\rm sp}^3$ hybridized bonding} with C, represented here by =, that the six atoms of a peptide group lie in a plane, and in fact the angles at C and N are nearly 120$^\circ$ as depicted.

This structure of backbone and peptide groups is conserved across all proteins.  This stripped-down description of a protein, which ignores residues, is thus comprised of a linear sequence of planar quadrilaterals, joined together at C$^\alpha$ vertices, where they meet at a tetrahedral angle as it turns out.  This part of protein geometry is thus highly constrained.

\begin{figure}%[tbhp]
\centering
\includegraphics[trim = 0 200 0 200,width=.75\linewidth]{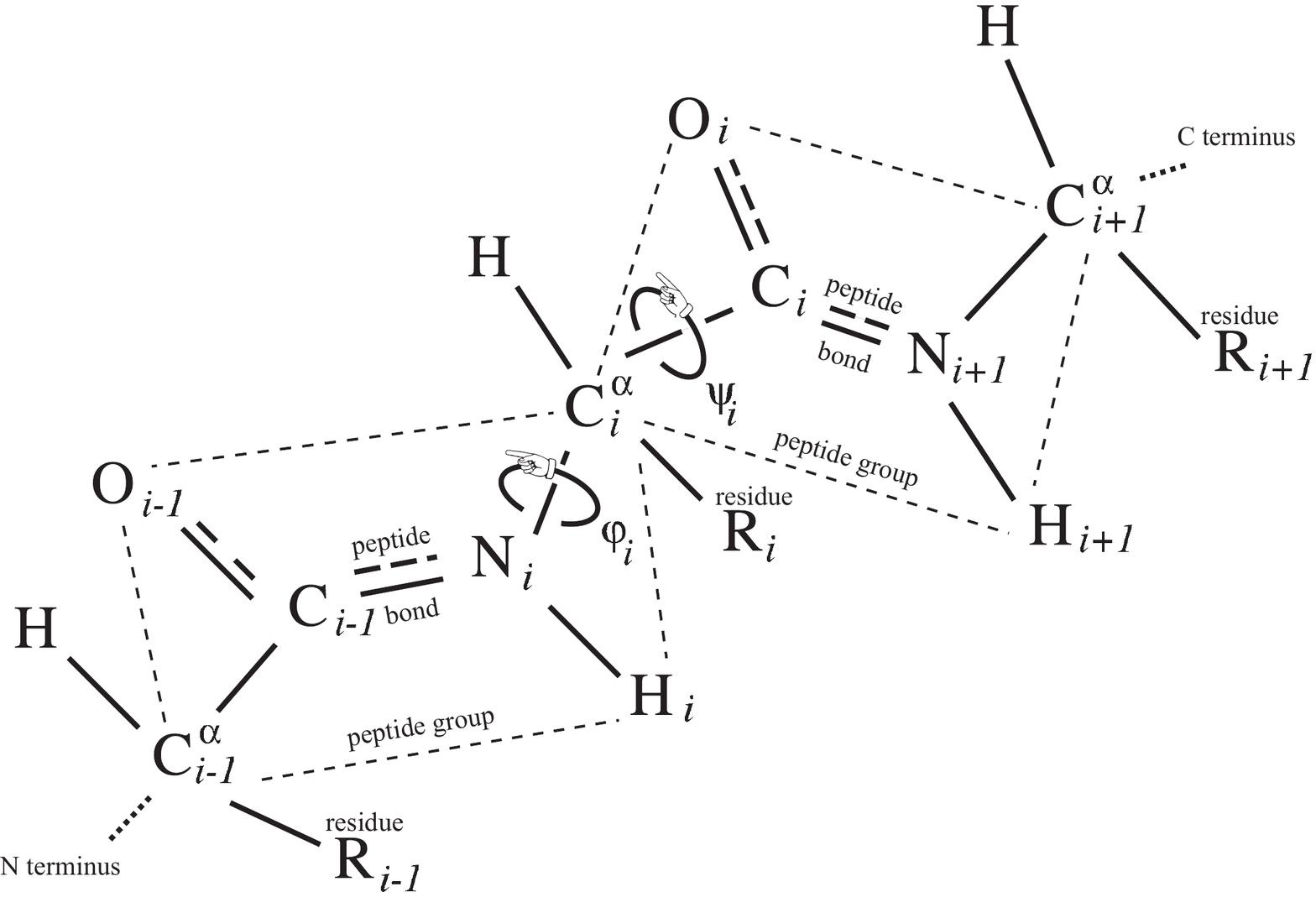}
\caption{ 
The conformational angles $\phi_i$ and $\psi_i$ between consecutive peptide groups
at the $i^{\rm th}$ residue of a protein.  Note parenthetically also the H comprising an amino acid attached to each C$^\alpha$, which was not mentioned in the main text.  This indexing of constituent atoms in a protein is called {\sl crystallographer's notation}. Figure adapted from 
[Penner, R. Moduli spaces and macromolecules.
{\em Bulletin of the American Mathematical Society}  {\bf 2016}, {\em 53}, 217--268].
}
\label{fig:conformational}
\end{figure}

The word of length from 3 to roughly 30,000, and typically of several hundred letters, in the (20+2+1)-letter alphabet of amino acid residues from N terminus to C terminus is called the {\sl primary structure} of the protein, and it uniquely determines protein identity, lacking however the decorations mentioned in the Introduction.
There are two essentially dihedral-angle type moduli illustrated in Fig.~1 for each-save-one amino acid residue comprising the protein, which are called the {\sl conformational angles}
$\phi_i$ and $\psi_i$ at the $i^{\rm th}$ residue.

Among the many forces that determine the protein conformation,
we shall concentrate here on the  {\sl hydrogen bonds}.  These occur when two electronegative atoms (meaning hungry for an electron, like O and N but not especially C) are nearby in space, and one of them covalently bonded to an H shares its electron cloud with the other hungry one.

\begin{figure}[!h]
\begin{center}
{{\epsfysize2.6in\epsffile{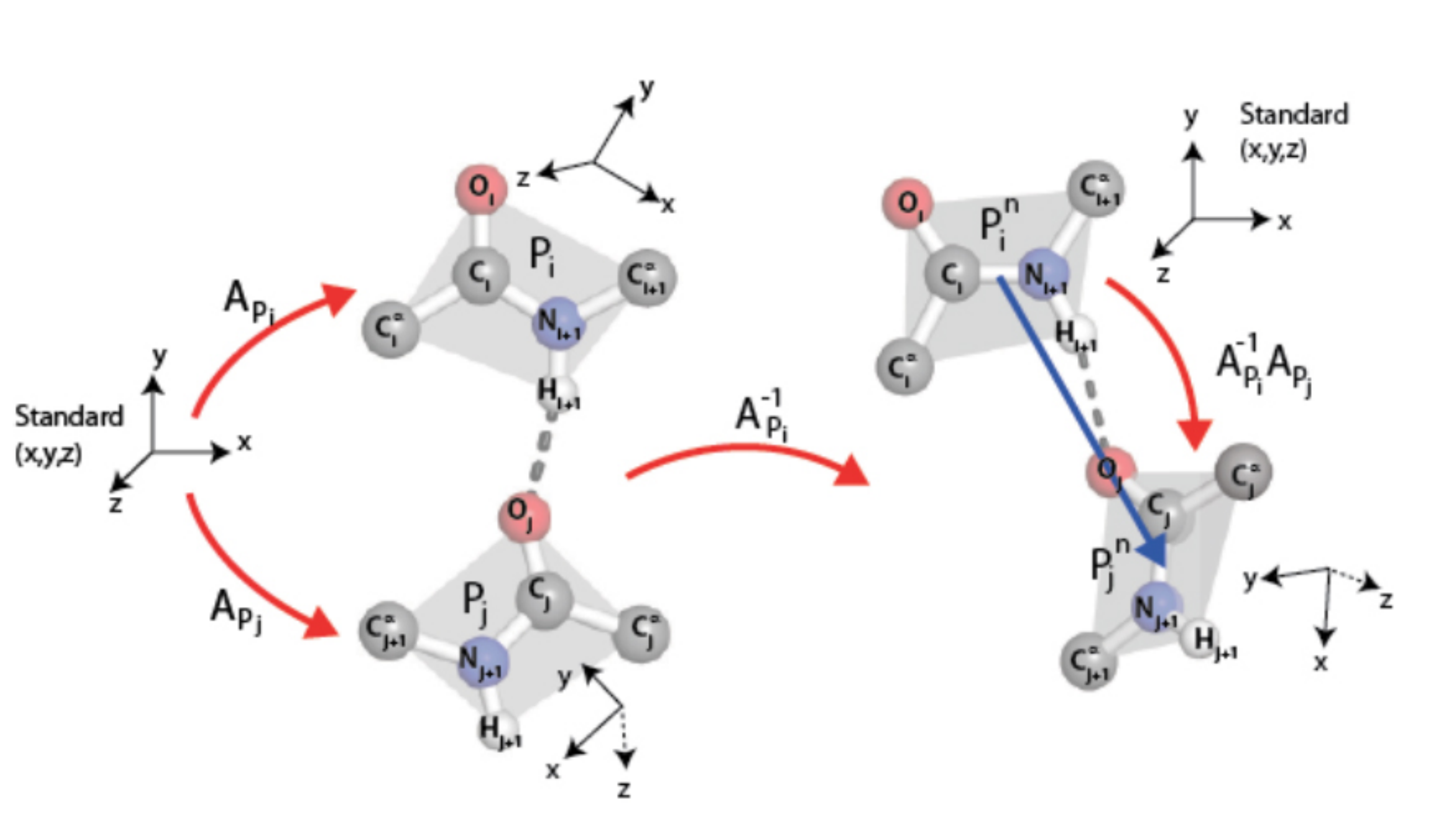}}}
\caption{Two peptide groups, illustrated in grey, are portrayed on the left participating in a BHB, depicted as a dashed line,   with donor ${\rm P}_i$ and acceptor
${\rm P}_j$. There is a unique ${\rm A}_{{\rm P}_i}\in{\rm SO(3)}$ carrying the oriented xz plane to the oriented plane of the peptide group for ${\rm P}_i$ and sending the positive x-axis to the ray of the peptide bond from ${\rm C}_i$ through ${\rm N}_{i + 1}$,
%\protect\overrightarrow{{\rm C}_i {\rm N}_{i + 1}}$, 
and likewise ${\rm A}_{{\rm P}_j}\in{\rm SO(3)}$ for ${\rm P}_j$. The composition
${\rm A} _{{\rm P}_i}^{-1} {\rm A}_{\small{\rm P}_j}^{}\in{\rm SO(3)}$ is the rotation associated to the pair ordered from donor ${\rm P}_i$ to acceptor ${\rm P}_j$.  Figure taken from \cite{pennerjcb1}.}
\label{fig:rotation}
\end{center}
\end{figure}

In the {\sl crystallographer's notation} of Fig.~\ref{fig:conformational},
where the backbone atoms ${\rm N}_i$ and ${\rm C}_i$ are adjacent to the $i^{\rm th}$ residue carbon ${\rm C}^\alpha_i$, imagine the protein backbone continuing in space and bringing a C$_j$=O$_j$ of one peptide group proximal to the H$_i$-N$_i$ of another peptide group as indicated in Fig.~\ref{fig:rotation}.  
They may form a hydrogen bond, to be denoted
C$_j$=O$_j$::H$_i$-N$_i$,  called a {\sl backbone hydrogen bond} or BHB with {\sl donor} N$_i$ and {\sl acceptor} O$_j$, which lie just a few Angstr\"om apart in space, though their necessarily non-zero {\sl distance $|i-j|$ along the backbone} may be large.

These BHBs taken together are said to form the {\sl secondary structure} of the protein.  
We say that the $i^{\rm th}$ residue ${\rm R}_i$ {\sl participates} in a BHB if either of its backbone-adjacent atoms ${\rm C}_i$ or ${\rm N}_i$ do.  Roughly 70\% of protein residues in a typical protein participate in BHBs. 
BHBs can also be bifurcated with one acceptor involved in two or even three different BHBs, though this higher quantum-chemical state is relatively uncommon.
Still other hydrogen bonds can form among residues, and between residues and C and N atoms in peptide groups, but we shall ignore them in the sequel and concentrate only on secondary 
structure\footnote{To interpret the standard {\sl cartoon representation} of protein introduced by Jane Richardson, let us mention the two
pervasive protein secondary structure motifs: the {\sl $\alpha$ helix} and the {\sl $\beta$ strand}.  An $\alpha$ helix is comprised of a consecutive sequence of N$_i$-H$_ i$ forming BHBs C$_{i+4}$=O$_{i+4}$::N$_i$-H$_ i$; $\alpha$ helices are nearly always right-handed and are represented in cartoons as helical ribbons.
In contrast, $\beta$ strands can be long-range along the backbone, where BHBs are formed between one sequence of consecutive backbone C and N atoms with another such sequence, either preserving
orientation ({\sl parallel}) of the two backbone segments or reversing it ({\sl anti-parallel}).  These strands,
represented as oriented ribbons, can combine into {\sl $\beta$ sheets}. }.

An essential database\footnote{You can access this database by typing ``RCSB" into your browser, query in English your desired protein such as ``hemoglobin" or ``SARS spike" at the top, click on a resulting PDB entry and then on ``1d-3d View" on the left for cartoon images.  Please do give it a try and start exploring this vast and marvelous resource.} in protein theory is the Protein Data Bank \cite{PDB} or PDB, which contains the 3-dimensional structures of nearly 200,000 proteins as of this writing.  The PDB provides the relative spatial coordinates in Angstr\"om of the centers of mass of each of the constituent atoms in the proteins catalogued in this database.  These 3-dimensional data  comprise the 
{\sl protein tertiary structure}\footnote{Some biologists deprecate the PDB as artificial since the proteins require chemical or other manipulation; moreover, the {\sl refinement} of experimental data to 3-dimensional structure is also arguably ad hoc. The former objection has been attenuated somewhat since recent years have witnessed a resolution revolution for Cryo Electron Microscopy methods, which can be less invasive. 
}.

\smallskip

\subsection{Geometry}\label{sec:geom}

As was described in the previous section, each peptide group lies in a plane due to quantum-chemical effects.   This plane comes equipped with an orientation, where the normal $\overrightarrow{{\rm C}^\alpha_i{\rm C}_i} \times
\overrightarrow{{\rm C}_i{\rm O}_i}$
to the $i^{\rm th}$ peptide group plane is given by the cross product $\times$ of displacement vectors in this plane in the notation of Fig.~\ref{fig:conformational}.  This oriented plane moreover contains the displacement vector $\overrightarrow{{\rm C}_i{\rm N}_{i+1}}$ of the peptide bond.

It follows that each peptide group determines a {\sl r\`epere mobile}, i.e., an ordered triple of pairwise orthogonal unit vectors so that the third is the cross product of the first and second in this order.  An ordered pair of such therefore determines a unique rotation of 3-space carrying the one to the other, that is, a point in the Lie group SO(3), which comes equipped with its bi-invariant metric from the Killing form and its Haar measure.  A binary relation on a collection of peptide groups therefore determines the subset of SO(3) given by its histogram.

We shall study the binary relation on peptide groups induced by BHBs, ordered from donor to acceptor.
Details are given in Fig.~\ref{fig:rotation}, in particular explaining the technical point, typical in the implementation of graph connections, that normalizing the r\'epere mobile of the donor peptide group to a standard position provides an element of SO(3) that depends only on the relative positions of the two constituent peptide groups in 3-space but not on their overall location.

%\vskip -0.4cm

We must determine a collection of peptide groups and BHBs whose histogram to compute, and there are two
aspects of this to discuss:  first, that the notion of hydrogen bond is not absolute, so we must choose a method
of recognizing them from a PDB file; and second, that the PDB itself has implicit biases, of 
{fashion}, of experimental facility, and with repeats of several identical monomers as frequently occur.  

For the former, there is a standard method of computing candidate hydrogen bonds, based on a crude energy estimation from the relative distances of constituent atoms, called the
Dictionary of Secondary Structures for Proteins (DSSP) \cite{dssp}.  To this we add
geometric constraints, as is often done, that in a DSSP-prospective BHB described in earlier notation by C=O::H-N, we impose the further stipulations:
HO distance $<$ 2.7\AA; NO distance $<$ 3.5\AA; $\angle$NHO$>$90$^\circ$; and $\angle$COH$>$90$^\circ$.

For the latter, one of several accepted methods of culling the PDB for unbiased representative subsets
is called PISCES \cite{pisces}.  This method determines a collection of PDB files for proteins whose primary structures are sufficiently dissimilar, yet are representative of the entire collection of primary structures in the PDB.  The degree of similarity of primary structure is described by a percentage of {\sl homology identity}, which we took to be 60\%, while not allowing monomer repeats.  There are also constraints of PDB file quality, and for completeness, we mention that we furthermore stipulated: the atomic resolution was $\leq$2\AA ~and the R factor $\leq$0.2, the latter of which describes the correlation between the measured data and the final tertiary structure of the PDB file.

Running PISCES on 12Mar2012 with these parameters produced a subset of the PDB; running DSSP using the parameters described on this set of PDB files yielded
a collection of 1166165 BHBs, and hence an equi-numerous collection ${\mathcal H}\subset{\rm SO(3)}$ of corresponding rotations.

In order to display an element of SO(3), recall from Euler that a rotation is uniquely determined by a unit vector $\vec u$ parallel to its axis of rotation together with its angle
$-\pi<\theta\leq\pi$ of rotation about that axis.  From this {\sl angle-axis} description, we derive a vector $\theta\vec u$ of length at most $\pi$, which we can plot in the ball of radius $\pi$ with antipodal points identified to give illustrations in a model of SO(3) $\approx{\mathbb R}{\rm P}^3$.

\begin{figure}[!h]
\begin{center}
{{\epsfysize2.6in\epsffile{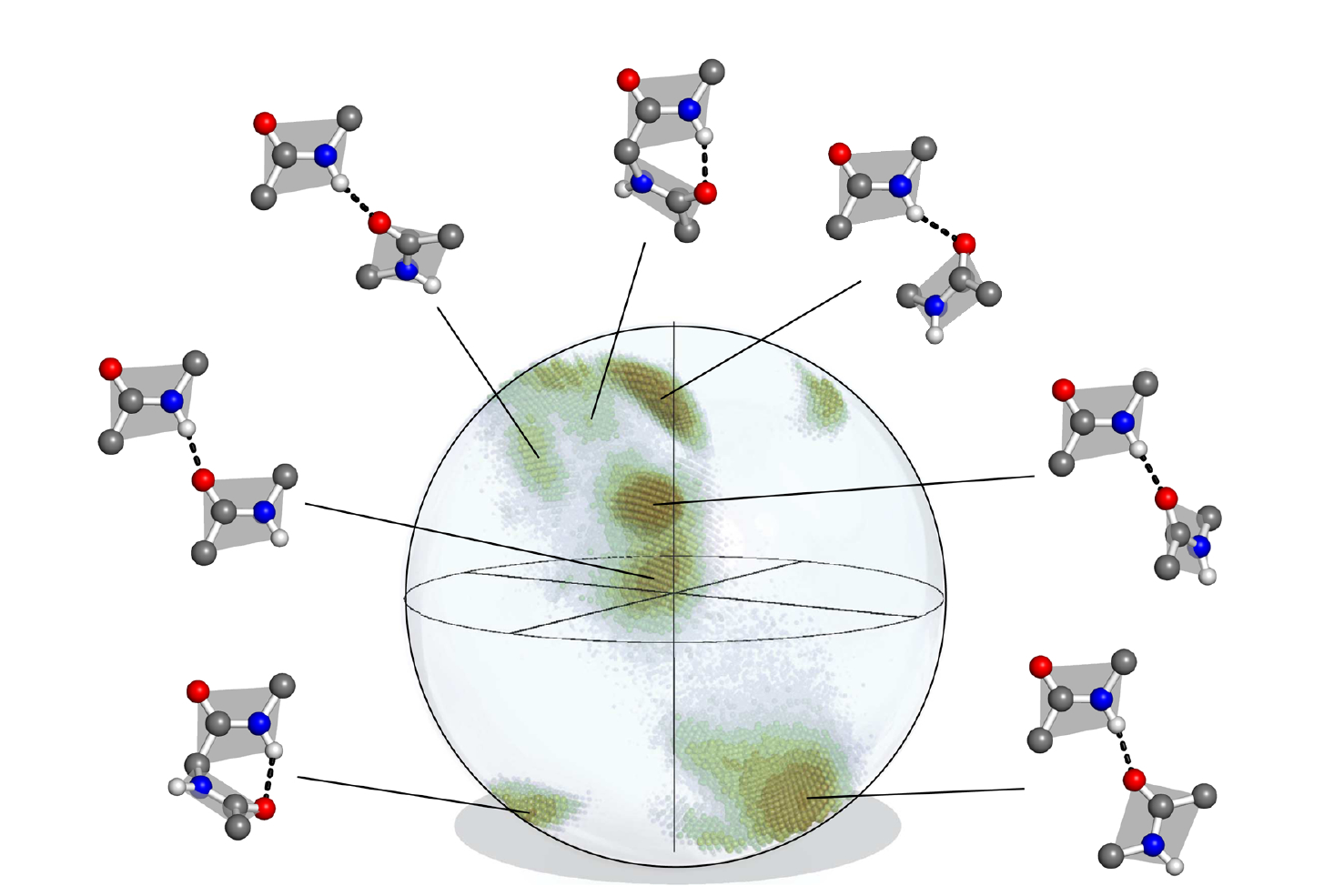}}}
\caption{Representation of ${\rm SO(3)}\approx {\mathbb R}{\rm P}^3$
as a ball of radius $\pi$ with antipodal points identified, containing the histogram ${\mathcal H}\subset {\rm SO(3)}$ of BHB rotations
for the subset of the PDB specified in the text.  Sample rotations between peptide groups for various BHBs, depicted by dashed lines, are also given. In order to shift the histogram away from the boundary sphere, an initial rotation is applied to the raw data determined as in Fig.~\ref{fig:rotation}; this normalizing rotation is given by left multiplication by the rotation $\theta=-2.479$, $\vec u=(-0.282, 0.907, -0.313)$ in angle-axis form.  A more detailed image of ${\mathcal H}$ is given Fig.~\ref{fig:slices}.}
\label{fig:graphabs}
\end{center}
\end{figure}

Displaying rotations in this manner produces the histogram ${\mathcal H}\subset{\rm SO(3)}$
depicted in Fig.~\ref{fig:graphabs}.  This distribution on SO(3) is at the heart of our methods,
and it is qualitatively robust under variations of the parameters used to compute it.

${\mathcal H}$ is contained in a region comprising only about 30\% of the volume of SO(3), though more than 95\% of the volume can be achieved by pairs of disjoint artificial abstract peptide groups at realistic displacements for hydrogen bonds.  As happens elsewhere, Nature is conservative.  Further details about ${\mathcal H}$ are provided in \cite{naturepaper}.

\subsection{Biophysics}\label{sec:phys}

There is a useful general principle in protein theory, the so-called {\sl quasi-Boltzmann Ansatz of Pohl-Finkelstein}, observed by Fritz Pohl \cite{pohl} and rigorously proved by Alexei Finkelstein and collaborators \cite{finkelstein1,finkelstein2}, on the {\sl free energy} $F$ of a protein detail, such as the rotation between planes of peptide groups participating in a BHB.
This Ansatz asserts that the occurrence of the detail is proportional to ${\rm exp} ({{-F}\over{kT_c}})$, where $k$ is the Boltzmann constant and $T_c$ is an effective temperature, called the {\sl conformational temperature}, which is quite near the melting temperature of the protein in degrees Kelvin.  

These are not Boltzmann statistics in the usual sense of a particle visiting states with a probability proportional to the energy divided by $- kT$, where $T$ is temperature, but rather reflect the statistics of words in the alphabet of amino acids that stabilize  the particular feature.

\begin{figure}[!h]
\begin{center}
{{\epsfysize3.9in\epsffile{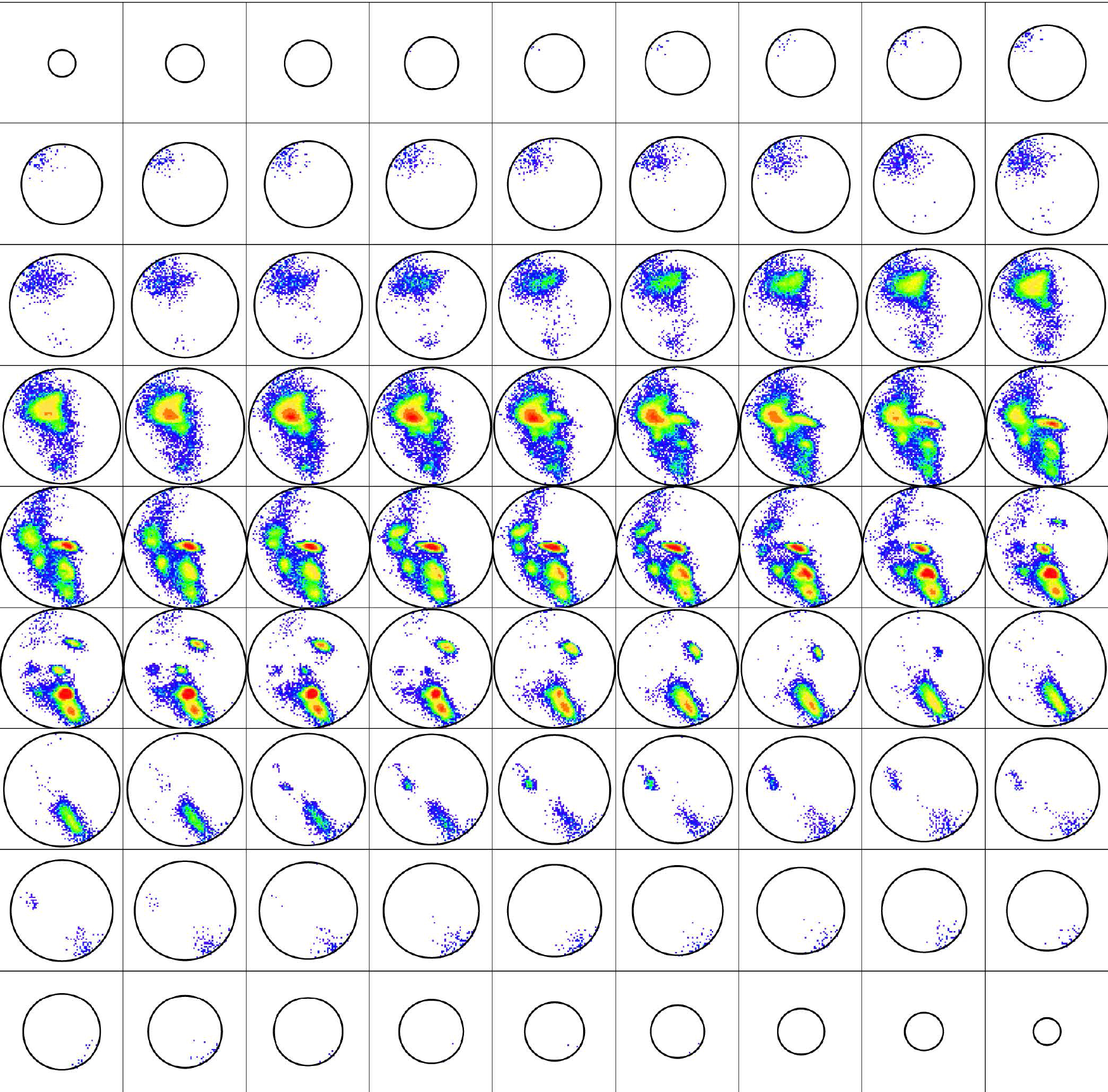}}}
\caption{Presented here are 81 horizontal slices of the histogram ${\mathcal H}\subset {\rm SO(3)}$ of BHB rotations from north to south pole colored by population density, where the R-Y-G-B color is linear in the density ranging from 19000 to 1.  The mode of $\mathcal H$ occurs for an internal turn of the ideal right-handed $\alpha$ helix in the fourth row from the top, fourth column from the left.  The ideal parallel (and anti-parallel, respectively) $\beta$ strand occurs in the sixth row from the top, first column from the left (and fifth row, between the fourth and fifth columns). Figure taken from \cite{naturepaper}.
}
\label{fig:slices}
\end{center}
\end{figure}

The qualitative meaning is that regions of low density in a distribution  represent protein details of high free energy and conversely.  Our practical consequence is that the histogram ${\mathcal H}\subset {\rm SO(3)}$ of BHB rotations determines a density on SO(3) in the natural way; namely, choose a decomposition of SO(3) into small {\sl boxes}, count members of ${\mathcal H}$ in each box, and divide by the Haar measure of the box to determine a piecewise-constant density $\rho:SO(3)\to{\mathbb R}$ which is constant on each box.  
A more refined illustration of this density than that given in Fig.~\ref{fig:graphabs} is displayed in Fig.~\ref{fig:slices}.  Details on boxes are given in \cite{naturepaper}.

%\vskip -1.in

It is important to emphasize that the distribution ${\mathcal H}\subset{\rm SO(3)}$ is computed empirically once and for all for the specified subset of the PDB and is well-defined up to left multiplication in SO(3).  The density $\rho:SO(3)\to{\mathbb R}$ is then canonically determined  by the choice of decomposition of SO(3) into boxes.  Given then another subject protein {\sl whose tertiary structure must be known}, its BHBs can be determined with the same specialization of DSSP and their rotations 
$P\subset {\rm SO(3)}$ 
computed and normalized, all just as discussed in the previous section.  

Given $p\in P\subset{\rm SO(3)}$, in keeping with the quasi-Boltzmann Ansatz, we define its {\sl relative backbone free energy} to be $$\Pi(p)={\rm log}_e[\rho(m)/\rho(p)],$$ where $m$ is the {\it mode}, i.e., point of highest density for $\rho$, which occurs for the ideal right $\alpha$ helix with $\rho(m)$ equal to 19000.  The box containing $m$ is saturated by $\alpha$ helices, and likewise for the boxes of the two ideal $\beta$ strands.  There are two {\sl affine ambiguities} in comparing these data:

\smallskip

$\bullet$ $\Pi(p)$ is given in units of $kT_C$, where $T_C$ is very nearly the
melting temperature, which varies between proteins.  Reasonable bounds for protein melting are
25$^\circ$-115$^\circ$ Celsius, or equivalently 298$^\circ$-388$^\circ$ Kelvin.  Since ${{388}\over{298}}\approx 1.3$, it introduces only marginal error, about which we shall comment further presently, to take a default conformational temperature of $T^*_C=350^\circ$ Kelvin to interpret all of the data, which gives $kT^*_C\approx 0.7$ kcal/mole; for comparison, the quantum of thermal fluctuation is $kT\approx 0.6~$kcal/mole at 300$^\circ$ Kelvin. This crude approximation resolves the {\sl homothetic ambiguity} from the underlying affine maps which are inherent in the quasi-Boltzmann Ansatz.

\smallskip
 
$\bullet$ It is typically only differences of free energy that are meaningful, essentially because the definition of entropy, cf. below, requires specification of macro-states, which may include the laboratory, 
continent, planet or universe where the experiments are conducted. 
There are theoretical computations in the protein literature for the free energy of the ideal helical turn, which is found to be -2 kcal/mole $\approx 2.9~kT^*_C$, and this can be used to advantage to fix the value at a point and resolve this {\sl translational ambiguity} of the underlying affine maps, using the saturation of boxes for ideal $\alpha$ helices and $\beta$ strands mentioned before.

\smallskip

Namely, the {\sl (absolute) backbone free energy} BFE of the rotation $p\in P$ is defined to be
$$[\Pi(p)-2.9]~kT_C^*\approx 0.7\times[\Pi(p)-2.9]~{\rm kcal/mole}.$$
This BFE can be directly compared for proteins with melting temperatures 
near 350$^\circ$ Kelvin, and even across different proteins with various melting
temperatures, with the understanding that it marginally underestimates for rotations in proteins
with higher melting temperatures and overestimates for lower ones. 

For a sense of
their distribution, approximate BFE values for rotations in ${\mathcal H}\subset{\rm SO(3)}$ followed in parentheses by their percentiles are: 
4.3 (50th), 5.6 (75th), 7.6 (90th), 8.5 (95th), 9.5 (99th) and 9.85 (100th).

The BFE of a residue $R$ in a protein is the maximum of the free energies
of the various BHBs in which it participates, namely
in crystallographer's notation, the BFE of $R_i$ is the maximum value of the BFEs
of the potentially several rotations of BHBs in which ${\rm C}_i$ and ${\rm N}_i$ participate.  
The BFE is undefined if there are no such BHBs.

\smallskip

But what physically is free energy $F$? 

\medskip

And where and why might it be useful in biology?  

\medskip

The (Helmholtz) definition $F=E-TS$, where $E,S,T$ are respectively energy, entropy, and temperature, rather avoids the question since it involves entropy, which is itself a slippery concept for most of us.  (This is discussed in some detail throughout \cite{AFbook}, to which we refer the reader for a more expert treatment.)  Minimizing free energy is tantamount to maximizing entropy at fixed temperature according to this equation, so free energy is minimized at equilibrium according to the Second Law of Thermodynamics.

More intuitively, the free energy is the energy of the system available for work, that is, it is {\sl free} or in other words {\sl available} to move things, while the {\sl entropic energy} $TS$ lingers in the ether, to wax poetic, out of reach for effective mechanical use.   

Specifically, in order that a protein stably maintain its characteristic {aperiodic crystalline} conformation necessary for biological function, its free energy overall must be negative.  On the other hand, in order for it to achieve a different biologically useful conformation absent other interactions, as with some proteins, there must be domains of positive free energy to drive the reconformation. These domains must be compensated by still other regions of low free energy to maintain stability.   

In the spirit of the quasi-Boltzmann Ansatz, rare protein details of any sort, necessary for work,  must be compensated by ordinary ones.  In the spirit of Charles Darwin, these high free energy regions, which have been selected by evolution, must have co-evolved with other low free energy regions to stabilize them.  One consequence is that domains of high free energy are likely to be of functional significance.

It follows that the utility of free energy in biology is pervasive.  Wherever there is concerted motion, there must be free energy of some sort to drive it, whether it be chemical, electrostatic, and so on, or in our case for BHBs, it is the BFE.  BHBs are transient in practice since they are just at the limits of stability in water, trading water entropy for hydrogen bond energy, as will be important in the next section.

\subsection{Biology}\label{sec:biol}

As we have discussed, the rare protein details necessary for work must be compensated by ordinary ones, and these high free energy regions, selected by evolution to serve some function, must have co-evolved with other low free energy regions that stabilize them.
There are important examples of such proteins with multiple conformations involving motility, signal transduction, pumping, and others, and most notably for the sequel certain viral processes as follows.  

Two sequential key events in the lifecycle of a virus, both of which are mediated by its glycoproteins, are its {\sl receptor binding} to an appropriate host cell
and its subsequent {\sl membrane fusion}\footnote{Fusion is sometimes with the cell membrane and sometimes with the membrane surrounding a vesicle called an {\sl endosome}, as is the case for SARS-CoV-2.  The endosome transports external material inside the cell, whether for disposal if threatening or for recycling if useful, and in either case is intended as housekeeping of the extracellular environment.  Viral fusion within an endosome is often driven by pH, since the endosomal pathway is in any case highly acidifying.} to access the cytoplasm of the cell.

Receptor binding must happen quickly in the chaotic 
environment of the host organism, its lungs, bloodstream or stomach for instance, all the while
guarding against the host immune defenses.  The ensuing fusion faces large energy barriers, derived
from removing the water between the two membranes, and involves tectonic conformational changes
in the glycoprotein.
Once bound and fused, the virus proceeds with its business of hijacking the native functions of the host to replicate itself and disperse to infect other cells, while still continuing to evade immune defenses.

In fact, BHBs are just at the limits of stability in water, and their consequent ephemeral nature is conducive to breaking and reforming, ideal for the quick transitions necessary for binding in the tumult of the host organism.  Once bound, viral glycoproteins behave like the transformer toy of a child, rapidly reconforming with subunits shifting by tens of Angstr\"om to be readied for fusion.  Viral glycoproteins are thus in fact metastable, just at the limits of stability, neither so stable as to defy such reconformation nor so unstable as to jeopardize reliable function, typically held in place by BHBs, among other forces.

It was therefore natural to study viral glycoproteins as a first test case for these BHB/BFE methods in \cite{pennerjcb1}.  Furthermore for several viruses, the same glycoprotein is represented in the PDB
both pre- and post-fusion, so the geometries of the two states could be compared directly.

What emerged from this analysis is that regions with large BFE predict large nearby conformational changes of the backbone, but not conversely.  Specifically, within one residue along the backbone of a residue whose BFE lies in the 90th percentile, there is a residue so that the sum $\phi+\psi$ of its conformational angles $\phi,\psi$ depicted in Fig.~\ref{fig:conformational} changes by at least 180$^\circ$.
One imagines a spring on a gate driving the gate to close, while there are other regions such as hinges with no particular signature of free energy but that concomitantly reconform.

With the pandemic underway, it was natural to apply these tools to coronaviruses in particular.  This was begun in \cite{pennerjcb2} to search for regions with large BFE in the spike glycoprotein files in the PDB, which were common to all human coronavirus diseases, including covid.  To functionally align glycoproteins from
different such viruses, a first homology alignment of the proteins provided neighborhoods along the backbone, with constituent residues then identified using similar BHB motifs of large BFE.  This technique is presumably of wider applicability.

Five such sites were found, and it was argued that these were promising so-called {\sl epitopes}, namely, targets for antibodies, both since their BFE suggested that interference would interrupt a crucial role in reconformation and since their conservation across different viruses suggested the unlikeliness of future mutation.

However, their unsuitability as specific epitopes for delivery through mRNA vaccines emerged from subsequent discussion with the vaccine development group at Moderna.  
%Once the mRNA has been taken up by the host cell and translated to protein, an important next step of the immune response is that the protein is diced into 10-15 residue long snippets by the Human Leucocyte Antigens (HLAs) and then transported and presented to the immune system at the cell membrane by the Major Histocompatibility Complex (MHC).  
A short stretch of high BFE residues evidently will not produce the same aperiodic crystal in isolation as it does in the full protein, precisely because it lacks the compensatory low BFE to stabilize it.  A high BFE snippet presented to the immune system will simply exhibit a structure different from its native geometry, so its presentation would not provoke the desired immune response. 

One might artificially stabilize a  high BFE snippet by chemical modification and utilize the modified snippet as mRNA vaccine cargo, akin to the current approach with its artificially stabilized full spike.
In lieu of this, one is led to search for conserved low BFE regions, suitable for 
%MHC-presentation if HLA-selected, 
presentation as epitopes to the immune system, which are adjacent to conserved high BFE regions, promising as mutation-resistant and function-impairing sites.  Several such regions were discovered in the SARS-CoV-2 spike, as reported in \cite{pennercmb}.  Other major obstacles to useful epitopes
%beyond HLA-selection 
remain, e.g., penetrating the glycogen coat on the spike.

This issue of carefully choosing putative epitopes for mRNA vaccine delivery is not simply academic.  We are seeing that the current approach, by both Pfizer/BioNTek and Moderna, of presenting the suitably stabilized full SARS-CoV-2 spike of older variants, has led to vaccine-escape by subsequent variants, especially by the current Omicron strains.  
Even with regulatory hurdles minimized and large-scale production streamlined, it seems we shall always be playing catch up with the virus.  

Moreover beyond immune-escape, this unsuitability of the full glycoprotein as a vaccine or therapeutic target becomes more severe for variants, or for other viruses, of higher morbidity.  A prospective approach based on targeting epitopes not prone to mutation and with higher specificity than the wild ones, rather than retrospectively presenting the full spike of past variants, might provide future advantage.

More recent work \cite{pennervac} turned attention specifically to mutations in the SARS-CoV-2 spike starting from the original ancestral strain called Wuhan-Hu-1. It was intriguing  to ponder how subsequent mutations might avoid the double-edged constraint from metastability in order to selectively mutate residues without increasing or decreasing BFE to respectively either disrupt stability or impede function.  

The first finding of \cite{pennervac} was that the mutation of a single residue participating in a BHB could impact the BFE profile along the entire spike molecule; satisfying the constraint of metastability in this case is thus achieved globally and not locally along the backbone.  This finding was based on the single mutation D614G, meaning that residue number 614 of the spike was changed from the amino acid D (short for Aspartic Acid) in Wuhan-Hu-1 to the amino acid G (short for Glycine) in the mutant.  This mutation occurred just after Wuhan-Hu-1 appeared and quickly globally overtook its ancestor, having improved the receptor binding abilities; one says in such a case that the mutation was {\sl selected} for its increase in {\sl fitness} of the virus.
%;spikes for both the ancestral Wuhan-Hu-1 and its D614G mutant appear in the PDB, as required for this analysis.

One imagines an underlying stochastic process of mutations in the genome, and hence of translated proteins\footnote{The {genetic code} discussed before mediates translation from genome to proteome, and there are biases in amino acid mutations resulting from the code itself.},  which are occasionally advantageous for fitness and therefore selected for propagation.  This is merely a formulation of Darwin's thesis in the context of viral mutation.

Taking together the so-called Variants of Concern, Variants of Interest and Variants under Monitoring 
before the advent of Omicron as a proxy for these selected mutations at the time, a clear pattern emerged in the second finding of \cite{pennervac}: By and large, the mutated residues in this collection of variants did not participate in BHBs in the Wuhan-Hu-1 ancestral variant\footnote{It is slightly more subtle.  The spike is comprised of two domains, S1 mediating binding and S2 mediating fusion.  Our finding about BHBs applies to S1 and not to S2.  Two likely explanations are first that S2 is active at much lower pH along the endosomal pathway, as discussed before, and second that S1 sits atop S2 fastening it in place before the two are cleaved by host proteins, also in the endosome.}.  Indeed, the simplest way to preserve delicate metastability is to avoid the constraints altogether by mutating residues not participating in BHBs, which do not themselves contribute to the BFE one way or another.  This elementary remark makes good sense.

Omicron challenges this paradigm in several regards.  First of all, the number of spike residues in Wuhan-Hu-1 selected for mutation in Omicron is much larger than among any of the other variants mentioned before, numbering in the thirties for the first Omicron strain called BA.1, as opposed to a just a few, namely, three to eight in each of the other strains, often shared by one or more of the selected variants.  Furthermore, most of these thirty or so residues which mutated in Omicron BA.1 do in fact participate in BHBs in Wuhan-Hu-1.

It is interesting to note that virtually all of these thirty or so residues are free from BHBs in the Delta variant, which was the predominant global strain pre-Omicron.  Furthermore
from the ancestral Omicron strain BA.1 to the currently predominant BA.2/BA.4/BA.5 strains of Omicron, all the nearly 30 further mutated spike residues\footnote{Most of these mutated residues occur in the {\sl N-terminal domain}, which is a prime location for epitopes, as is consistent  with the proclivity of Omicron variants for immune-escape.} save one (residue number 547) do not participate in BHBs in the highest quality PDB file for BA.1 (the PDB file with accession number 7WP9).   This strongly confirms the earlier finding.

There are two related explanations \cite{ferret} for these anomalies of Omicron: the virus may have mutated in an immunocompromised individual,  or it may have been sequestered in a small isolated group of hosts before exiting to the general population only when it was a fit competitor to other variants.
In each case, the relief from the pressure to compete--with the compromised  host immune system in the former case and with other variants at large in the latter--removes constraints on the genome and permits greater flexibility to vary in mutation. The phylogenetic tree for SARS-CoV-2 \cite{nextstrain} shows the early divergence of Omicron from the Wuhan-Hu-1 strain without much spread in infected population until later, lending support to the second explanation.  

Our finding that residues free from BHBs are more likely to mutate than those participating in BHBs has the corollary that in order to preserve the residues selected for fitness, it is favorable for them to be protected from mutation by participating in BHBs.  On the other hand, those residues that are not critical for fitness should have their BHBs selectively removed by mutation to allow for their subsequent mutation so as to potentially increase fitness.  

This process of removing BHBs from Wuhuan-Hu-1 to allow mutation in Omicron presumably occurred during its latent period of circulation in a small isolated group, or perhaps in the evolution to Delta, as proposed in \cite{pennervac}, which was then contracted by an immunocompromised individual. 
In the latter case, the phylogenetic data could reflect {\sl convergent mutation}, which is the propensity for fitness-improving mutations to occur more than once in a population. 

The main theoretical conclusion of \cite{pennervac} is stated thus: {\sl Protein BHBs provide a regulatory network governing viral mutation in the spike glycoprotein of SARS-CoV-2, conserving those BHBs that are selected for fitness and removing those that are not.} One might postulate this as a more sweeping principle for glycoproteins of other RNA viruses, or even beyond\footnote{We extrapolate from mutation of the one viral glycoprotein to others, and possibly still more generally, in the spirit of Jacques Monod's memorable remark that ``What is true for E. Coli is true for the elephant," which pertains despite the First Law of Mathematical Biology in the Introduction.}.

Mutation takes place on the level of the genome and not directly on the level of the proteome.  These are linked through the selective pressure for improving fitness, which is implicit in our analysis of the data by considering only the successful variants mentioned before.  These have evidently been so selected as evidenced by their propagation. 

There is an amusing further aspect of our postulate as follows.  Just as amino acid residues in proteins participate in BHBs, so too do the nucleic acid bases of RNA participate in Watson-Crick bonds\footnote{Just as there are other hydrogen bonds in protein beyond BHBs, there are also other {\sl non-canonical bonds} between bases in RNA beyond those of Watson-Crick, all of which (including Watson-Crick) are comprised of hydrogen bonds.}, together termed the {\sl secondary structure} of the RNA molecule. 

It is an easy consequence of Claude Shannon's information theory that {\sl if you slow down, then you will make  fewer mistakes}.  Studies show both that 
transcription/polymerization  is retarded by secondary structure of the nascent RNA\footnote{This also protects from deleterious secondary structure which is long-range along the RNA sugar-phosphate backbone from forming during transcription.} and that translation is retarded by secondary structure of the mature mRNA, both of which should therefore improve fidelity. Furthermore, RNA bases buried in secondary structure are less exposed and hence protected from chemical degradation in the cell, such as {\sl deamination},
which can also cause unselected mutation. 

In contrast to our empirically motivated postulate on BHBs, this is a conceptual argument, based on polymerase, ribosomes\footnote{In fact, multiple ribosomes, called {\sl polysomes}, travel simultaneously along the mRNA.  These can collide without secondary structure to impede them and thereby degrade the mRNA. Repair of the degraded mRNA potentially introduces accidental mutations not selected for fitness, giving yet another sense in which mRNA secondary structure protects the genome from potentially unfavorable mutation.
In any case, the degraded mRNA
has inferior productive longevity, and consequently Moderna untranslates the spike to mRNA cargo to maximize secondary structure.}  and degradation, that mRNA secondary structure protects participating regions from mutation.  Mirroring the logic for proteins, it is therefore salutary to protect critical selected bases by preserving the secondary structure in which they participate and deleting it  elsewhere for subsequent potentially fitness-increasing mutation.

Combined with our postulate on proteins, it follows that
regions in viral mRNA participating in nucleic acid secondary structure code for regions in protein participating in amino acid secondary structure and conversely, at least for RNA viruses, or at any rate at least for their receptor binding domains.  This is an arguably speculative  {\sl principle of biophysical mirror symmetry} on secondary structures for protein and viral mRNA\footnote{Though the arguments linking secondary structure to polymerase fidelity and degradation avoidance may apply more generally, this biophysical mirror symmetry principle pertains only to mRNA and not to other RNAs, and only for RNA viruses.  We do not discount the possibility that the principle may hold more generally, though it does not take account of mutational errors during DNA replication.}.

Offering a potential partial explanation and refinement, the theorized RNA world posits that terrestrial life originated with RNA performing all functions, and subsequently developed to the world in which we exist with its several interacting bio-macromolecules.  In that RNA world, the proto-protein functions of RNA could reasonably be expected to be reflected in the proto-protein geometry of RNA, going beyond secondary even to the tertiary structure of protein being mirrored in RNA, albeit not necessarily in what would become mRNA, likely also including non-canonical bonds, and at least for proteins in ancient organisms, many of which
provide avatars for contemporary proteins.

Offering a potential confirmation is difficult directly due the relative paucity\footnote{The principal repository for nucleic acid structures is the Nucleic Acids Database (NDB).  The NDB is derived from the PDB with additional search capabilities specialized to nucleic acids.  The NDB catalogues about 1,650 RNA structures as of this writing.} of known 3-dimensional mRNA structures, especially the dearth of sequences long enough to reflect native structure.
However, predictions of canonical RNA secondary structure can be based on its primary structure of sequential nucleic acids and the rules of Watson-Crick pairings,
cf. \cite {Huang} and the references therein.  Since our catalogue\footnote{There are several libraries of nucleic acid primary structures, shared together under the umbrella International Nucleotide Sequence Database Collaboration.}
 of RNA primary structures is vast, this presents a possible avenue for testing the biophysical mirror symmetry principle.

\end{document}